\documentclass{elsart}
\usepackage{amsmath}
\usepackage{epsf}
\usepackage{graphicx}

\journal{Astroparticle Physics}

\begin{document}

\begin{frontmatter}

\title{Photon Gas Thermodynamics in Doubly Special Relativity}

\author[pku]{Xinyu Zhang},
\author[pku]{Lijing Shao},
\author[pku,chep,chps]{Bo-Qiang Ma\corauthref{cor}}\ead{mabq@pku.edu.cn}
\address[pku]{School of Physics and State Key Laboratory of Nuclear Physics and Technology,
\\Peking University, Beijing 100871, China}
\address[chep]{Center for High Energy Physics, Peking University, Beijing
100871, China} %
\address[chps]{Center for History and Philosophy of Science, Peking
University, Beijing 100871, China}%
\corauth[cor]{Corresponding author.}

\begin{abstract}
Doubly special relativity (DSR), with both an invariant velocity and
an invariant length scale, elegantly preserves the principle of
relativity between moving observers, and appears as a promising
candidate of the quantum theory of gravity. We study the
modifications of photon gas thermodynamics in the framework of DSR
with an invariant length $|\lambda|$, after properly taking into
account the effects of modified dispersion relation, upper bounded
energy-momentum space, and deformed integration measure. We show
that with a positive $\lambda$, the grand partition function, the
energy density, the specific heat, the entropy, and the pressure are
smaller than those of special relativity (SR), while the velocity of
photons and the ratio of pressure to energy are larger. In contrast,
with a negative $\lambda$, the quantum gravity effects show up in
the opposite direction. However, these effects only manifest
themselves significantly when the temperature is larger than
$10^{-3} E_{\rm P}$. Thus, DSR can have considerable influence on
the early universe in cosmological study.
\end{abstract}


\end{frontmatter}

\clearpage

\section{Introduction}\label{sec1}
Some developments of quantum gravity (QG) suggest a smallest length
scale for the structure of space-time, or equivalently, an upper
energy bound for particles in the quantum geometrical
background~\cite{AmelinoCamelia:2000ge,AmelinoCamelia:2000mn,KowalskiGlikman:2001gp,KowalskiGlikman:2001ct,Magueijo:2001cr,Magueijo:2002xx,Magueijo:2002am,AmelinoCamelia:2002vy,AmelinoCamelia:2010pd,AmelinoCamelia:2005ik,Salesi:2009kd}.
The most natural candidate for the minimal length appears to be the
Planck length $l_{\rm P} \equiv \sqrt{G\hbar/c^3} \simeq 1.6 \times
10^{-35}$~m (or correspondingly, the Planck energy, $E_{\rm P}
\equiv \sqrt{\hbar c^5/G} \simeq 1.22 \times 10^{19}$~GeV, for the
maximal energy of particles). These typical constants can arise from
the combination of the quantum ($\hbar$), the relativity ($c$), and
the gravity ($G$).\footnote{However, there are also arguments that a
new fundamental scale might appear rather than the conventional
Planck scale, see \textit{e.g.}, Ref.~\cite{Shao:2010vj} and
references therein.} In such scenarios, the Lorentz symmetry of
space-time breaks down at quantum-gravitational scale, and this
might leave ``relic'' effects at relatively lower energies and
modify low energy physics with extra
terms~\cite{Mattingly:2005re,AmelinoCamelia:2008qg,Liberati:2009pf,Xiao:2010yx}.

However, there arises an apparent puzzle --- it is well-known that
in the special relativity (SR), the length of an object transforms
between two observers with relative movements according to
Lorentz-Fitzgerald contraction, so in whose reference frame is the
smallest length scale which is mentioned above measured? This
problem is deeply related to an essential property of physical laws
among inertial frames, the so-called principle of relativity, which
is regarded as a milestone in the progress of physics. However, in
the domain of quantum gravity physics, it is not obvious that this
principle still holds firmly. Instead, it deserves the most careful
contemplations.

Amelino-Camelia~\textit{et al.} suggested a way to reconcile the
paradox between the relativity and the minimal
length~\cite{AmelinoCamelia:2000ge,AmelinoCamelia:2000mn,KowalskiGlikman:2001gp,KowalskiGlikman:2001ct,Magueijo:2001cr,Magueijo:2002xx,Magueijo:2002am,AmelinoCamelia:2002vy,AmelinoCamelia:2010pd,AmelinoCamelia:2005ik,Salesi:2009kd}.
There are two constants that are preserved in such a theory, so it
is usually called ``doubly special relativity''
(DSR)~\cite{AmelinoCamelia:2000mn,AmelinoCamelia:2010pd}. DSR
preserves the relativity between inertial frames, whereas deforms
the Lorentz algebra with nonlinear actions as a cost. Consequently,
the well-known dispersion relation for a particle in SR, $E^2 = p^2
+ m^2$, has to be modified in the form with extra QG imprints,
\begin{equation}
E^2 = p^2 + m^2 + \eta E^3 + ... \,,
\end{equation}
where $\eta$ is a parameter believed to be of the order of the
Planck length. Meanwhile the laws of transforming energy and momenta
between different inertial observers are also explicitly deformed in
DSR. Hence it is possible that a single energy or momentum scale is
invariant. We should emphasize that modified Lorentz transformations
in momentum space can have physically substantial effects, even
though the modifications for different kinds of particles are
identical, since the departure from SR can manifest its effects when
we measure and compare physical observables in different conditions
(for detailed discussions, see \textit{e.g.},
Ref.~\cite{AmelinoCamelia:2010pd}).

It is interesting that such modifications lead to many further
predictions which can be tested in a series of experiments. One
serious example is the energy-dependence of light speed, and its
effect is relevant to $\gamma$-ray burst
observations~\cite{AmelinoCamelia:1997gz,AmelinoCamelia:9697,Ellis:9900,ellis01,Xiao:2009xe,Shao:2009bv,Shao:2010wk},
some corrections to the predictions of inflationary
cosmology~\cite{Alexander:2001dr} and dark
energy~\cite{MersiniHoughton:2001su}. Indeed, although some
theoretical calculations motivate the modification of the dispersion
relation, the fact that a modified dispersion relation is
experimentally measurable by itself is a more exciting reason to
take it into serious consideration.

As pointed out in
Refs.~\cite{KowalskiGlikman:2001ct,AmelinoCamelia:1999pm,AmelinoCamelia:2009tv},
the law of composition of momenta in SR has to be modified together
with the modification of the dispersion relation, since the theory
is relativistic with two invariant quantities, and accordingly the
role of integration over energy-momentum space is modified. There
are some studies~\cite{Camacho:2007qy,Das:2010gk} on the effects
from the modified dispersion relation on a photon gas. In our paper,
we study the modified thermodynamics of the photon gas in the
framework of DSR, with more careful considerations compared to
previous studies. We properly include the effects of the modified
dispersion relation, the deformed integration measure, and the upper
energy-momentum bound. We find that the number of available
microstates is modified and consequently thermodynamical quantities
are altered in DSR. We show that with a positive parameter
$\lambda$, the grand partition function, the energy density, the
specific heat, the entropy, and the pressure of the photon gas are
smaller than those of SR, while the velocity of photons and the
ratio of pressure to energy are larger. In contrast, with a negative
$\lambda$, the quantum gravity effects show up in the opposite
direction. However, these effects only manifest themselves
significantly when the temperature is larger than $10^{-3} E_{\rm
P}$.

The paper is organized as follows. In Sec.~\ref{sec2}, we review the
procedure to get the dispersion relation for the photon gas in the
framework of DSR proposed and generalized by Magueijo and
Smolin~\cite{Magueijo:2001cr,Magueijo:2002xx,Magueijo:2002am}. In
Sec.~\ref{sec3}, we derive the modified integration measure and the
grand partition function of the photon gas. In Sec.~\ref{sec4}, we
study various thermodynamical quantities of the photon gas in
details, and the cases $\lambda>0$ and $\lambda<0$ are compared with
those of SR ($\lambda=0$). In Sec.~\ref{sec5}, we provide summaries
of the paper.

\section{Modified Dispersion Relation For Photon Gas}\label{sec2}
Theoretically, DSR itself cannot decide the dispersion relation used
to describe the real nature, whereas only experiments can pick out
the right
formula~\cite{AmelinoCamelia:2000mn,AmelinoCamelia:2010pd}.
Unfortunately, the current status of observations is incapable to
draw a decisive conclusion. However, promising observations are
emerging~\cite{Mattingly:2005re,AmelinoCamelia:2008qg,Liberati:2009pf,Xiao:2009xe,Shao:2009bv,Shao:2010wk}
and many possibilities to modify the dispersion relation in the
framework of DSR have already been proposed.

In this paper, the model we adopt is DSR2, which was proposed and
generalized by Magueijo and
Smolin~\cite{Magueijo:2001cr,Magueijo:2002xx,Magueijo:2002am}. This
model introduces a nonlinear modification to the action of the
generators of Lorentz group in momentum space, and this renders the
theory to be compatible with an observer-independent length scale.
The modified boost generators are given
as~\cite{Magueijo:2001cr,Magueijo:2002xx,Magueijo:2002am}
\begin{equation}\label{U}
K^i = U^{-1} [p_0] L_0^{\ i} U [p_0],
\end{equation}
where $L_{ab} = p_a {\partial}/{\partial p^b} -
 p_b {\partial}/{\partial p^a}$ are the standard
generators of Lorentz group, and $U[p_0]$ is defined as
\begin{equation}
U[p_0](p_a)=\frac{p_a}{1-\lambda p_0},
\end{equation}
with the parameter $\lambda$ believed to be around the Planck length
scale. From the above construction, with $U[p_0]$ defined to map the
energy-momentum manifold onto itself,
\begin{equation}
U \circ (E, \vec{p})=(f_1 E,f_2 \vec{p}),
\end{equation}
any isotropic dispersion relation can be written in the following
form~\cite{Magueijo:2001cr,Magueijo:2002xx,Magueijo:2002am},
\begin{equation}\label{dispersion relation}
E^2 f_1^2(E,\lambda)-p^2 f_2^2(E,\lambda) = m^2.
\end{equation}

In this paper, we consider the case with $f_1 = 1$ and $f_2 =
1+\lambda E$. The dispersion relation for the photon with zero mass
becomes
\begin{equation}\label{dr}
p = \frac{E}{1+\lambda E}.
\end{equation}
When $\lambda>0$, Eq.~(\ref{dr}) implies a maximum momentum $p_{max}
= \lambda^{-1}$, which is an invariant under deformed transformation
laws. In contrast, $\lambda<0$ corresponds to an energy upper bound
for photons. Contrary to the modified SR in the framework of
effective field theory where the dispersion relation for photons is
spin-dependent, the dispersion relation is spin-independent in DSR.

\begin{figure}
\includegraphics[width=0.9\textwidth]{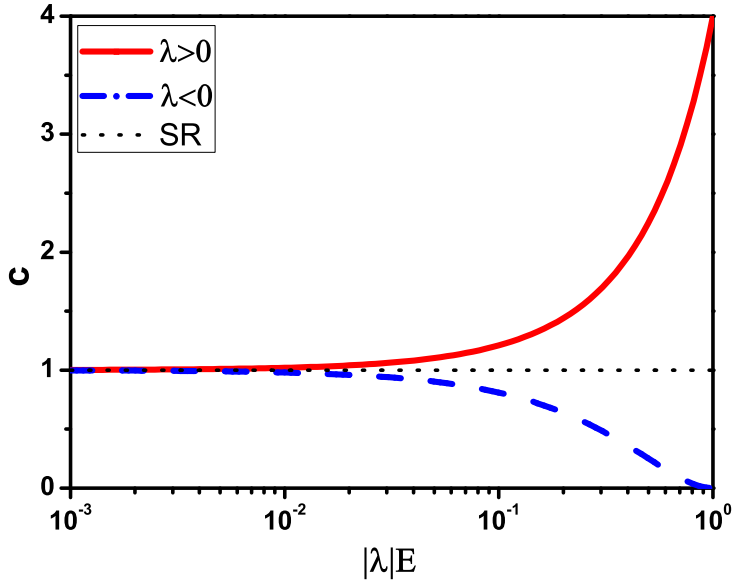}
\caption{\small The relation between the speed of light and the
energy of photons. The red solid and blue dashed lines are results
for scenarios where $\lambda>0$ and $\lambda<0$, respectively. The
dotted horizontal line is the canonical result in the
SR.}\label{speedoflight}
\end{figure}

Based on the relation (\ref{dr}), one obtains
\begin{equation}\label{ddr}
\frac{\mathrm{d}p}{\mathrm{d}E} = \frac{1}{(1+\lambda E)^2}.
\end{equation}
The speed of light is therefore energy-dependent,
\begin{equation}
c(E) = \frac{\mathrm{d}E}{\mathrm{d}p} = (1+\lambda E)^2,
\end{equation}
which is shown in Fig.~\ref{speedoflight}. We can see explicitly
that with energy increasing, the speed of light increases with a
positive $\lambda$, and decreases with a negative one. If the theory
is to provide a solution to the horizon problem without inflation,
$\lambda>0$ is required. However, we consider both signs in this
paper, for the sake of completeness.

We have to emphasize that, as stressed by
Amelino-Camelia~\cite{AmelinoCamelia:2000ge,AmelinoCamelia:2000mn,AmelinoCamelia:2010pd},
though many authors working on DSR insist on getting formulas that
make sense all the way up to infinite particle energies, it is
possible that DSR is relevant only at energy scales that are
sub-Planckian (but nearly Planckian). If we actually expect to give
up an intelligible picture of space-time and its symmetries above
the Planck scale, then perhaps we should be open to the possibility
of using mathematics that provides an acceptable (closed) logical
picture of DSR only at the leading order (or some finite orders) in
the Taylor expansion of formulas in powers of the Planck length.
This possibility will not affect the results in most literatures,
where only the effects at low energies were anaylzed. However, in
this paper, the integration in the phase space is up to $E_{\rm P}$,
so we need to assume that the formulas are tenable even up to the
Planck scale in order to be self-consistent.

\section{Density Of Microstates and Grand Partition Function}\label{sec3}
There are possibilities of photon decays, {\it e.g.}, $\gamma
\rightarrow e^{+} + e^{-}$, when the dispersion relation
(\ref{dispersion relation}) alone is considered with proper LV
parameters and unmodified energy-momentum conservation
laws~\cite{Mattingly:2005re,Shao:2010wk}. However, these
possibilities are totally invalid in DSR, since such kind of decays
lead to energy thresholds for the decay of a single particle, and
those are forbidden for the sake of relativistic
principle~\cite{AmelinoCamelia:2010pd}. Therefore it is proper to
consider the photon gas composed from free particles forming a
thermodynamical ensemble.

Consider the photon gas in a container of volume $V$. We assume that
the momentum spectrum is continuous in the thermodynamical limit
with an infinite $V$ and negligible boundary conditions. Then the
number of microstates available in the position ranging from
$\vec{r}$ to $\vec{r}+\mathrm{d}\vec{r}$ and momentum ranging from
$\vec{p}$ to $\vec{p}+\mathrm{d}\vec{p}$ is given by
\begin{equation}
\mathrm{d} N = 2\times\frac{1}{(2\pi\hbar)^3}
\mathrm{d}\vec{r}\mathrm{d}\vec{p},
\end{equation}
where the factor $2$ results from two directions of polarization.

\subsection{Derivation Of Modified Integration Measure}
The arguments reported in
Refs.~\cite{KowalskiGlikman:2001ct,AmelinoCamelia:1999pm,AmelinoCamelia:2009tv}
imply that the net effect of a deformed measure of integration in
DSR can be described with the replacement,
\begin{equation}\label{deformedintegration}
\mathrm{d}^4 p \rightarrow \theta(E)\mathrm{d}^4 p,
\end{equation}
where $\theta(E)$ is a function of the energy $E$ and needs to be
adjusted by future insights on detailed analysis.
Eq.~(\ref{deformedintegration}) is often left out without proper
justifications in previous works~\cite{Das:2010gk}, and we would
consider the deformation of the integration measure for
completeness.

For the dispersion relation we adopt in (\ref{dr}), the law of
composition of momenta is modified into
\begin{equation}
p \oplus \Delta p =
\left(E\left(\vec{p}\right)+E\left(\Delta\vec{p}\right), \vec{p} +
\theta(E)\Delta\vec{p}\right),
\end{equation}
where $\Delta p$ is an infinitesimal increment. Because of the
virtue of relativistic principle, $p \oplus \Delta p$ satisfies the
same functional form of Eq.~(\ref{dr}). Consequently, it is
straightforward to get the form of $\theta(E)$,
\begin{equation}
\theta(E) = \frac{1}{(1+\lambda E)^2}.
\end{equation}
Thus for each spatial component $p_i$ and a function
$F^{\prime}(p_i)$, which is the integrand of $F(p_i) = \int
F^{\prime}(p_i) \mathrm{d}p_i$, the law of composition of momenta
suggests that,
\begin{equation}
F^{\prime}(p_i) = \lim_{\Delta p_i\rightarrow 0} \frac{F(p_i \oplus
\Delta p_i)-F(p_i)}{\Delta p_i} = \frac{\partial F(p_i)}{\partial
p_i} \frac{1}{(1+\lambda E)^2}.
\end{equation}
This in turn suggests that for one spatial momentum we have
\begin{equation}
F = \int F^{\prime}(p_i) \mathrm{d}p_i = \int \frac{\partial
F(p_i)}{\partial p_i} \frac{1}{(1+\lambda E)^2}\mathrm{d}p_i,
\end{equation}
which is equal to a replacement of the integration measure,
$\mathrm{d} p_i \rightarrow (1+\lambda E)^{-2} \mathrm{d} p_i$. In
the case we are interested, with three spatial and one time
dimensions, we finally get the modified integration measure,
\begin{equation}\label{deformed3}
\mathrm{d}^4 p \rightarrow \frac{1}{(1+\lambda E)^6} \mathrm{d}^4 p.
\end{equation}

It is worthy to mention that the form of the deformed measure of
integration (\ref{deformed3}) is a little different from that used
in
Refs.~\cite{KowalskiGlikman:2001ct,AmelinoCamelia:1999pm,AmelinoCamelia:2009tv}.
Their results can be written in a united form as
\begin{equation}
\mathrm{d}^4 p \rightarrow \e^{-3 \eta E/E_p} \mathrm{d}^4 p.
\end{equation}

If we assume that the modified dispersion relations adopted in our
work and of earlier papers coincide in the leading-order
approximation, we have
\begin{equation}
\lambda \sim \frac{\eta}{2 E_p},
\end{equation}
and consequently,
\begin{equation}
\e^{-3 \eta E/E_p} - \frac{1}{(1+\lambda E)^6} = \e^{-6 \lambda E}
-\frac{1}{\left(1+\lambda E \right)^6}  \sim \mathcal
{O}\left((\lambda E)^2\right).
\end{equation}

We should emphasize that high-order differences for the integration
measure result from different modified dispersion relations and
mathematical frameworks. Nowadays, these differences do not really
matter, because currently we cannot conclude which specific model of
DSR should describe the real nature. Nevertheless, the leading-order
corrections, which are of great scientific interests at the present
stage, are the same in different models of DSR.

\subsection{Derivation Of Modified Grand Partition Function}

After we carefully consider the integration measure, we now turn to
the grand partition function of a photon gas. The grand partition
function $\Xi$ of the photon gas in a container with fixed volume
$V$ is defined as
\begin{equation}
\ln \Xi = - \int \frac{8 \pi V}{(2\pi\hbar)^3} p^2
\ln(1-e^{-\frac{E}{k_B T}}) \frac{1}{(1+\lambda E)^6} \mathrm{d}p.
\end{equation}
In the following we will always adopt the units in which
$\hbar=k_B=1$.

In the scenario where $\lambda>0$, with relation between $E$ and $p$
given in Eqs.~(\ref{dr}) and (\ref{ddr}), the grand partition
function is given in the form
\begin{equation}
\ln \Xi = - \int_ 0^{\infty} \frac{8 \pi V}{(2 \pi
)^3}\frac{E^2}{(1+\lambda E)^{10}}\ln\left(1 -
e^{-\frac{E}{T}}\right)\mathrm{d}E.\label{xi}
\end{equation}
We adopt a dimensionless variant $x=E/T$ to simplify (\ref{xi}),
\begin{equation}
\ln \Xi = -\frac{V T^3}{\pi ^2} \int_ 0^{\infty} \frac{x^2
\ln(1-e^{-x})}{(1+\lambda T x)^{10}}\mathrm{d}x.
\end{equation}
In the limit $\lambda\rightarrow 0$, it reduces to the normal SR
result,
\begin{equation}
\ln \Xi = \frac{\pi^2}{45} V T^3.
\end{equation}

The grand partition function $\Xi$ of the photon gas when
$\lambda<0$ can be obtained similarly, and the result is
\begin{equation}
\ln \Xi = -\frac{V T^3}{\pi ^2} \int_ 0^{-\frac{1}{\lambda T}}
\frac{x^2 \ln\left(1-e^{-x}\right)}{(1+\lambda T x)^{10} }
\mathrm{d}x.
\end{equation}
The change of the upper limit in the integration results from the
existence of a maximum energy $-1/{(\lambda T)}$. Still, it reduces
to the normal SR result in the limit $\lambda\rightarrow 0^{-}$. We
would like to point out that in this case, the integration diverges
when $T$ approaches to 1, and this implies that underlying
quantum-gravitational theories should replace the DSR model when the
temperature is around or larger than the Planck temperature. We
observe that most models, in which an invariant energy or momentum
scale exists with the speed of light decreasing with respect to
energy, suffer from similar problems. In spite of this problem, we
can mainly pay attention to the situation where $T$ is lower bounded
from 1, thus the theory is still well defined.

For both $\lambda>0$ and $\lambda<0$, we further set $\lambda=\pm 1$
respectively, and that is only a matter of units without losing
generality. We write the grand partition function $\Xi$ in the
following form,
\begin{equation}
\ln \Xi = \frac{\pi^2}{45} V T^3 f(T),
\end{equation}
with
\begin{equation}
f(T)= -\frac{45}{\pi^4}\int_ 0^{\infty}\frac{x^2
\ln(1-e^{-x})}{(1+xT)^{10}}\mathrm{d}x,~~~~\lambda=1,
\end{equation}
and
\begin{equation}
f(T)= -\frac{45}{\pi^4}\int_ 0^{\frac{1}{T}} \frac{x^2
\ln\left(1-e^{-x}\right)}{(1- T x)^{10} }
\mathrm{d}x,~~~~\lambda=-1,
\end{equation}
for compact discussions below. Thus, all modifications from DSR are
encoded in the function $f(T)$.

\begin{figure}
\includegraphics[width=0.9\textwidth]{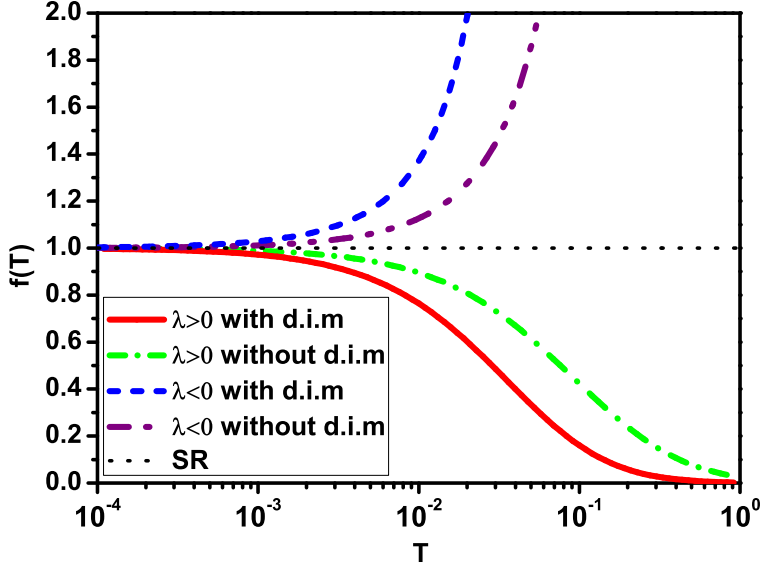}
\caption{\small The behavior of $f(T)$. The red solid line and the
blue dashed line are the results of the situations in which
$\lambda=1$ and $\lambda=-1$ respectively. The green dash-dotted
line and the purple dash-dot-dotted line are corresponding results
without taking the deformed integral measure (d.i.m) into
consideration. The dotted horizontal line is the canonical result in
SR.}\label{ft}
\end{figure}

We show the behavior of $f(T)$ in Fig.~\ref{ft}. It is noteworthy to
observe that, $f(T)<1$ with $\lambda>0$ while $f(T)>1$ with
$\lambda<0$. Thus the number of available states changes according
to $\lambda$. When $\lambda$ is positive, the number of available
states allowed is reduced, and while $\lambda$ is negative, it is
enhanced. This is the main result we get from thermodynamics of DSR.
From Fig.~\ref{ft}, it is illustrated that the deformation of
integration measure makes the changes more sharply. It should be
carefully taken into account especially when $T$ is large. At lower
temperature $T \ll 1$, all lines meet the SR case with $f(T)=1$.

\section{Thermodynamic Properties}\label{sec4}

Having derived the expression of the partition function of the
photon gas, we can study various thermodynamical properties in the
following. Both the situations in which $\lambda>0$ and $\lambda<0$
are to be considered and compared, and we call them ``positive-type
theory" and ``negative-type theory" respectively. As will be shown,
all the thermodynamical variants depend merely on $f(T)$ and its
derivatives. Furthermore, we have checked that the results presented
here are almost the same even when some other forms of DSR models
are considered instead of DSR2. Thus, our results are possible to
represent general deformed dispersion relations qualitatively within
DSR theories.

\subsection{Internal Energy}

\begin{figure}
\includegraphics[width=0.9\textwidth]{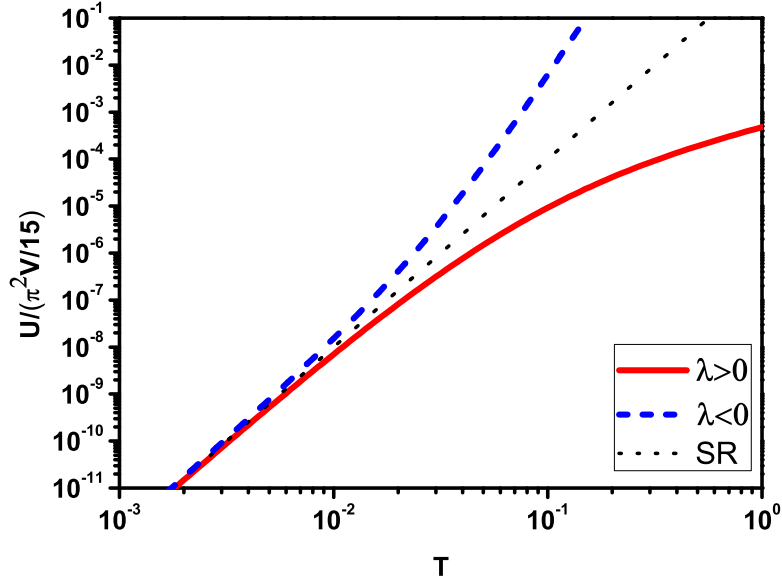}
\caption{\small The behavior of $U(T)$. The red solid line and the
blue dashed line are the results of the situation in which
$\lambda=1$ and $\lambda=-1$ respectively. The dotted horizontal
line is the canonical result in SR.}\label{ut}
\end{figure}

The expression of the internal energy $U$ for the photon gas is
given by
\begin{equation}
U = T^2 \frac{\partial \ln \Xi}{\partial T} = \frac{\pi^2 V T^4}{15}
(f+\frac{T}{3} f^{\prime}),
\end{equation}
where $f^{\prime}=\frac{\mathrm{d}f(T)}{\mathrm{d}T}$. Now the
internal energy depends on both $f$ and $f^\prime$. We plot the
internal energy $U$ versus $T$ in Fig.~\ref{ut}. It is shown that,
compared to the case in SR, the internal energy decreases in the
positive-type theory, and increases in the negative-type theory, and
this can be roughly understood through the number of available
states from Fig.~\ref{ft}. And the DSR results seem to be compatible
with SR results when $T<0.01$.

\subsection{Specific Heat}

\begin{figure}
\includegraphics[width=0.9\textwidth]{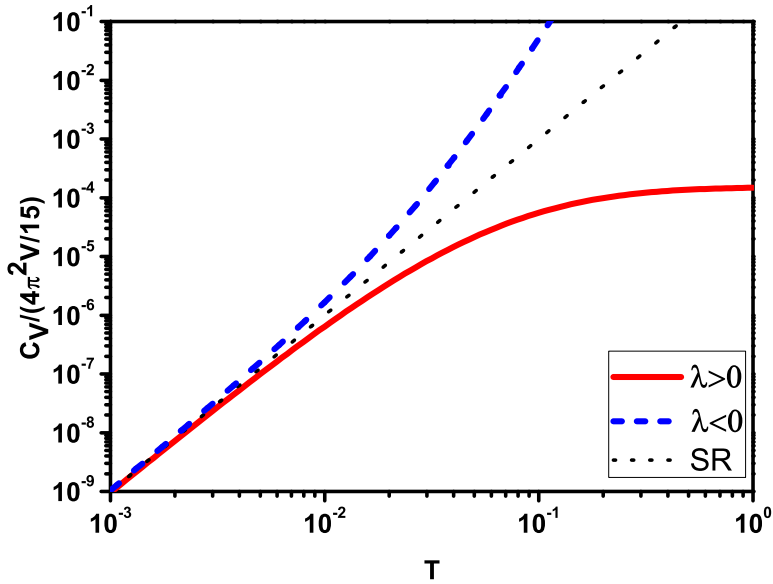}
\caption{\small The behavior of $C_V(T)$. The red solid line and the
blue dashed line are the results of the situation in which
$\lambda=1$ and $\lambda=-1$ respectively. The dotted horizontal
line is the canonical result in SR.}\label{cv}
\end{figure}

The specific heat of the photon gas $C_V$ is given as
\begin{equation}
C_V = \left(\frac{\partial U}{\partial T}\right)_V = \frac{4 \pi^2 V
T^3}{15} (f+\frac{2T}{3} f^{\prime}+\frac{T^2}{12}
f^{\prime\prime}),
\end{equation}
which now depends on $f^{\prime\prime}$ besides $f$ and $f^\prime$.
We show the results in Fig.~\ref{cv}. Compared to the result in SR,
the specific heat is remarkably smaller in the positive-type theory,
and larger in the negative-type theory, when $T$ approaches to $1$.
It is interesting to notice that in the positive-type theory, the
specific heat tends to be a constant when $T$ is large. The smaller
$C_V$ and larger $C_V$ than that of SR in the positive-type and
negative-type theories respectively reflect the situations in
Fig.~\ref{ut}, where the energy increases slowly in positive-type
theory and diverges in negative-type theory.

\subsection{Entropy}

\begin{figure}
\includegraphics[width=0.9\textwidth]{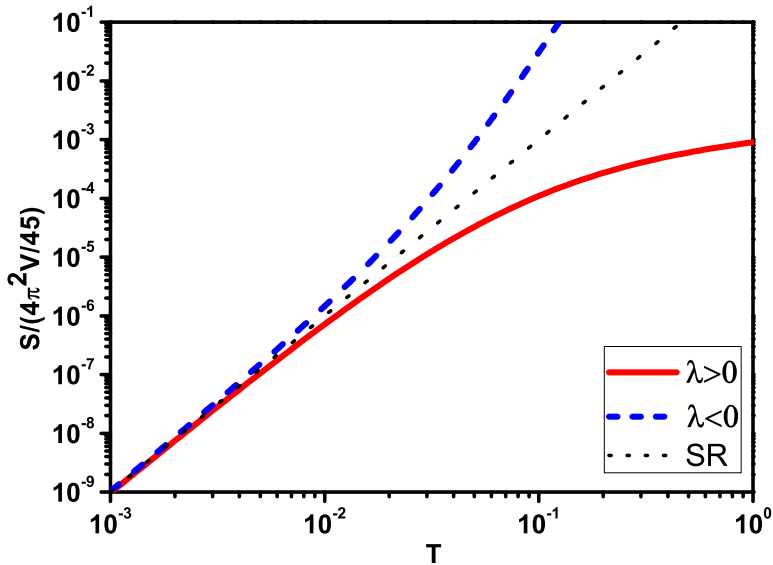}
\caption{\small The behavior of $S(T)$. The red solid line and the
blue dashed line are the results of the situation in which
$\lambda=1$ and $\lambda=-1$ respectively. The dotted horizontal
line is the canonical result in SR.}\label{st}
\end{figure}

We can also obtain the interesting thermodynamical quantity, entropy
$S$,
\begin{equation}
S = \left(\ln \Xi + T \frac{\partial \ln \Xi}{\partial T}\right) =
\frac{4 \pi^2 V T^3}{45} \left(f +\frac{T}{4}f^{\prime}\right),
\end{equation}
which now, like the internal energy $U$, depends on $f$ and
$f^\prime$. The result of the relation between entropy and
temperature is illustrated in Fig.~\ref{st}. It is clearly seen that
the entropy grows much slower in the positive-type theory and faster
in the negative-type theory, and this is consistent with our
previous observations. In fact, the different behaviors of the
entropy of the system for two types of theories result directly from
different modifications of the total available number of
microstates. When $\lambda>0$, there are fewer available states,
hence less chaos, while $\lambda<0$ induces more available states
and more chaos. The entropy measures chaos naturally. Thus the
behavior of the entropy is rather understandable.

\subsection{Pressure and Pressure-Energy Density Relation}

\label{secprho}

\begin{figure}
\includegraphics[width=0.9\textwidth]{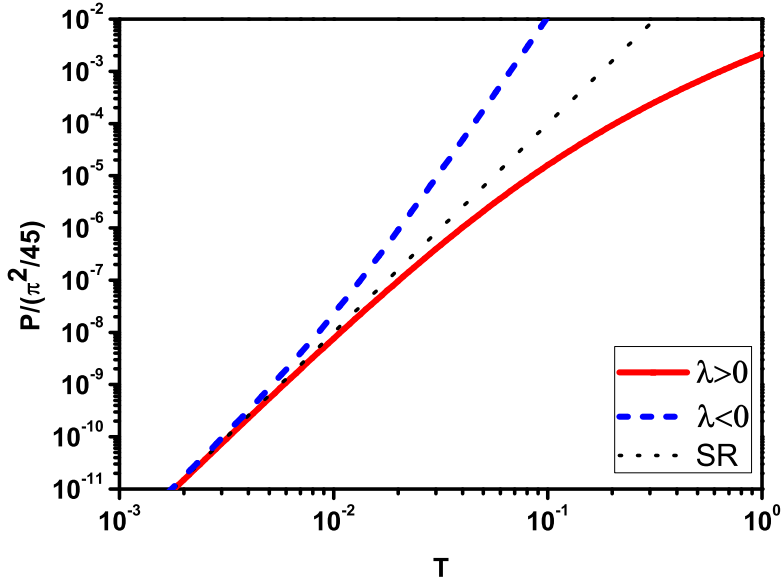}
\caption{\small The behavior of $P(T)$. The red solid line and the
blue dashed line are the results of the situation in which
$\lambda=1$ and $\lambda=-1$ respectively. The dotted horizontal
line is the canonical result in SR.}\label{pt}
\end{figure}

The expression of $\Xi$ can be used to get the pressure $P$ of the
photon gas
\begin{equation}
P = T \frac{\partial \ln \Xi}{\partial V} = \frac{\pi^2}{45} T^4
f(T),
\end{equation}
which appears merely a function of $f$. The relation between the
pressure $P$ and $T$ is given in Fig.~\ref{pt}. We can see that the
pressure reduces when $\lambda>0$, with respect to the case in SR,
and reverses when $\lambda<0$. It is understandable that the fewer
number of states ($\lambda>0$) causes the photons to gather with an
effective attractive force, while the more number of density induces
an effective repulsive force, compared to the case in SR. Such an
effective force is important when the dynamical behavior of photons
is under consideration.

\begin{figure}
\includegraphics[width=0.9\textwidth]{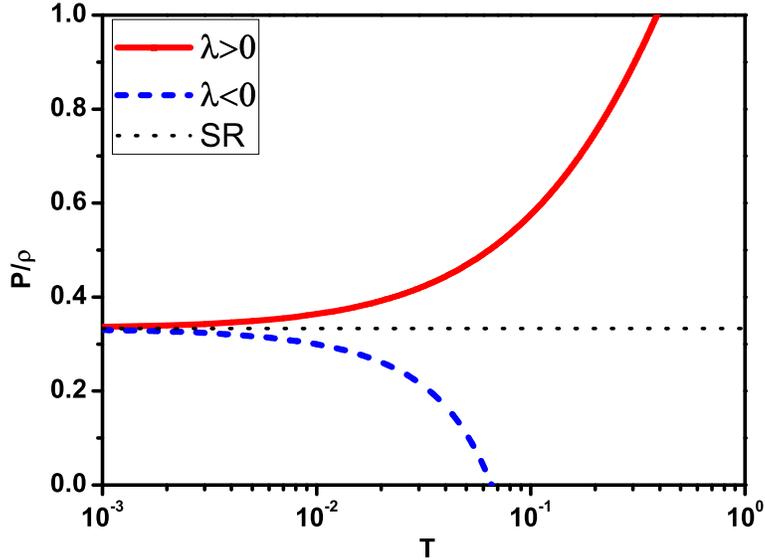}
\caption{\small The behavior of ${P}/{\rho}$. The red solid line and
the blue dashed line are the results of the situation in which
$\lambda=1$ and $\lambda=-1$ respectively. The dotted horizontal
line is the canonical result in SR.}\label{prho}
\end{figure}

Although one cannot directly detect the internal energy $U$ of the
photon gas, the relation between the pressure $P$ and the energy
density $\rho = U/V$ is measurable experimentally and is considered
to be important to various aspects. The modifications of $P/\rho$,
in principle, are effects appearing at the macroscopical level and
can have significant influences on the early universe. We derive the
relation,
\begin{equation}
\frac{P}{\rho}=\frac{1}{3+T\frac{f^{\prime}}{f}},
\end{equation}
whose modification now depends on the ratio of $f^\prime$ and $f$.
We plot $P/\rho$ via $T$ in Fig.~\ref{prho}. It is clearly seen that
in the ultra-relativistic regime for photons, the well-known
canonical relation, $P = \rho/3$, has to be modified. In contrast
with our naive expectations, $P/\rho$ with a positive $\lambda$ is
larger than that with a negative one. It is caused by the negative
slope of $f(T)$ when $\lambda>0$, and the positive slope of $f(T)$
when $\lambda<0$.

\section{Conclusion}\label{sec5}

In this paper, we derived the grand partition function of a photon
gas in doubly special relativity (DSR), based on careful
considerations of the deformed dispersion relation and the modified
measure of integration, as well as the upper bound of
energy/momentum. Then we discussed the thermodynamical quantities of
the photon gas and showed that the behaviors of thermodynamical
variants are modified according to the deformation parameter
$\lambda$.

These modifications are not remarkable when the temperature is low,
say $T<10^{-3} E_{\rm P}$, thus it is not easy to detect the effects
directly in present laboratory experiments. However, significant
differences exist when the energy approaches to the Planck scale.
Therefore, it is suggested that these modifications play an
important role in cosmology, especially on properties of the early
universe. Thus if these modifications turn out to be correct, the
behavior of photons in the early stage of the Big Bang will be very
different from what we observe at low energy. For example, the
modification of the ratio of pressure to energy density we have
considered in Sec.~\ref{secprho} leads to a different characteristic
of the evolvement of the weight that radiations occupy among all
types of energy in the whole universe, and its consequence is of
noticeable importance on the evolution of the universe.

In summary, we derived the thermodynamics of the photon gas in the
framework of DSR. We carefully included effects from the modified
energy-momentum dispersion relation, the deformed integration
measure, and the upper bound of energy/momentum, and some of them
are often left out by other studies. It is shown in detail that
different behaviors, other than those in special relativity, emerge
when the energy approaches to the quantum gravity scale. Thus, the
results could have significant consequences on the early universe,
such as the inflation, in cosmological study.

\section*{Acknowledgments}

This work is supported by National Natural Science Foundation of
China (Nos. 11021092, 10975003, 11035003, and 11005018), and
National Fund for Fostering Talents of Basic Science (Nos. J0630311,
J0730316). It is also supported by Principal Fund for Undergraduate
Research at Peking University.

\end{document}